\documentclass[%
 aip,
 amsmath,amssymb,
 reprint,%
]{revtex4-1}

\usepackage{graphicx}
\usepackage{dcolumn}
\usepackage{bm}
\usepackage{siunitx}
\usepackage{xcolor, soul} 

\usepackage[utf8]{inputenc}
\usepackage[T1]{fontenc}
\usepackage{mathptmx}

\begin{document}

\preprint{AIP/123-QED}

\title[Acoustic Metal Particle Focusing in a Round Glass Capillary]{Acoustic Metal Particle Focusing in a Round Glass Capillary}

\author{M. S. Gerlt}
 \email{gerlt@imes.mavt.ethz.ch}
\author{A. Paeckel}
\author{A. Pavlic}
\affiliation{ 
Mechanics and Experimental Dynamics, Department of Mechanical and Process Engineering, ETH Zurich.
}
\author{P. Rohner}
\author{D. Poulikakos}
\affiliation{ 
Laboratory of Thermodynamics in Emerging Technologies, Department of Mechanical and Process Engineering, ETH Zurich.
}
\author{J. Dual}
\affiliation{
Mechanics and Experimental Dynamics, Department of Mechanical and Process Engineering, ETH Zurich.
}

\date{\today}

\begin{abstract}
Two-dimensional metal particle focusing is an essential task for various fabrication processes. While acoustofluidic devices can manipulate particles in two dimensions, the production of these devices often demands a cleanroom environment. Therefore, acoustically excited glass capillaries present a cheap alternative to labour-intensive cleanroom production.
Here, we present 2D metal micro-particle focusing in a round glass capillary using bulk acoustic waves. Excitation of the piezoelectric transducer at specific frequencies leads to mode shapes in the round capillary, concentrating particles towards the capillary centre.
We experimentally investigate the particle linewidth for different particle materials and concentrations. We demonstrate the focus of copper particles with $\sim \SI{1}{\micro\meter}$ in diameter down to a line of width $60.8 \pm \SI{7.0}{\micro\metre}$ and height $45.2 \pm \SI{9.3}{\micro\metre}$, corresponding to a local concentration of $\SI{4.5}{\percent}$ v/v, which is $90$ times higher than the concentration of the initial solution. 
Through numerical analysis, we could obtain further insights into the particle manipulation mechanism inside the capillary and predict the particle trajectories. We found that a transition of the acoustic streaming pattern enables us to manipulate particles close to the critical particle radius.
Finally, we used our method to eject copper particles through a tapered round capillary with an opening of $\SI{25}{\micro\metre}$ in diameter, which would not be possible without particle focusing.
Our novel setup can be utilized for various applications, that otherwise might suffer from abrasion, clogging and limited resolution.
\end{abstract}

\maketitle
\section{Introduction}\label{sec:metal_intro}
Acoustophoresis, the manipulation of particles utilising acoustic forces, is one of the most popular techniques for particle manipulation because it is non-invasive, label-free, and biocompatible. \cite{10.1039/9781849737067} To a great extent, acoustofluidic devices are based on microchannels fabricated in silicon or polydimethylsiloxane (PDMS).\cite{10.1038/s41378-019-0064-3} Despite the flexibility in design, the production of silicon microchannels relies on expensive and labour-intensive cleanroom procedures and, depending on the resolution, PDMS microchannels require silicon stamps produced in a cleanroom environment.\cite{10.1146/annurev.matsci.28.1.153} Further, acoustic particle manipulation inside PDMS cavities usually requires interdigital transducers (IDTs), which need to be designed carefully and evaporated with high-end equipment.

An approach involving less expensive equipment and reduces manual labour is the application of glass capillaries in combination with bulk acoustic waves. Here, a piezoelectric element (piezo) is glued to an off the shelf glass capillary. On excitation of the piezo with an AC signal, the capillary can vibrate in specific modes that lead to beneficial acoustic potentials inside the capillary. This procedure has been used for various applications such as biomedical analysis,\cite{10.1063/1.1376412} blood trapping,\cite{10.1063/1.4863645} seed particle trapping for sample washing,\cite{10.1039/c2lc40697g} nanoparticle enrichment, \cite{10.1021/acsnano.6b06784} and two dimensional concentration of microparticles. \cite{10.1063/1.5142670} Additionally, intensive numerical investigations have been carried out concerning acoustic particle manipulation inside glass capillaries. \cite{10.3390/mi11030240, 10.1121/1.4754547, 10.1103/PhysRevApplied.8.024020}

In this publication, we show, for the first time, two dimensional focusing of metal particles that are close to the acoustofluidics' theoretical radiation force driven manipulation size limit in common rectangular microfluidic channels.\cite{10.1039/C2LC21261G} We characterised our device's performance by analysing the particle linewidth for different particle sizes, materials, flow rates and concentrations. Our thorough numerical investigations regarding capillary displacement, Gor'kov potential, and acoustic streaming reveal a prominent resonance frequency of the capillary close to $\SI{1.75}{\mega\hertz}$. Further, our simulations disclose a local minimum in the average streaming velocity while maintaining a high acoustic radiation force, explaining our ability to focus particles to a narrow line.

Finally, we employed our setup for the stable ejection of highly concentrated metal particles using a nozzle with an opening diameter of $\SI{25}{\micro\metre}$. Given that the nozzle's aperture is only $\sim 23$ times bigger than the particle diameter, a continuous ejection without clogging would not be feasible without sufficient particle focusing.

Our investigations and novel setup are relevant for a wide variety of industrial applications that rely on particle ejection out of small nozzles, including, e.g., metal 3D printing and waterjet cutting.

\section{Operating Principle}\label{sec:metal_op}
\begin{figure*}[t!]
\centering
  \includegraphics[width=0.8\linewidth]{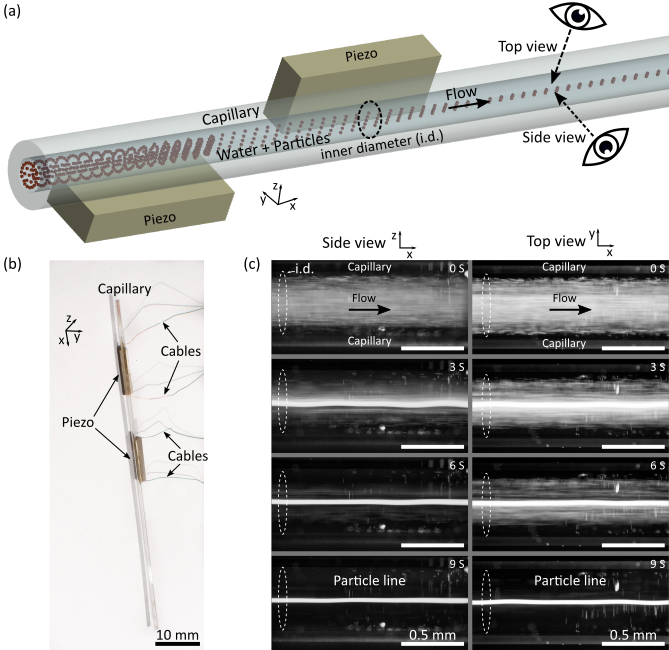}
  \caption{\textbf{Design and working principle of two dimensional particle focusing in a round glass capillary.} (a) Sketch of the design. (b) Photograph of the acoustofluidic device with a 10 mm scale bar. (c) Micrograph series of two dimensional focusing with $\SI{5}{\micro\metre}$ diameter fluorescent polystyrene particles from side and top view. The particles were pumped through the device at a flow rate of $\SI{100}{\micro\liter\per\minute}$. The piezoelectric elements were excited at $\SI{1.67}{\mega\hertz}$ with $\SI{20}{\volt}_\mathrm{PP}$. The initial polystyrene concentration was $\SI{0.5}{\percent}$ volume to volume ratio (v/v). Scale bars corresponding to $\SI{0.5}{\milli\metre}$
}
  \label{metal.fgr:op}
\end{figure*}

The acoustofluidic device consists of a round glass capillary with two piezoelectric elements (piezos) attached to it (Fig. \ref{metal.fgr:op} (a)). Upon excitation of the piezos with frequency $f$, the capillary starts to vibrate. When a suitable resonance of the fluid cavity inside the capillary is excited, particles migrate towards the centre of the capillary, forming a thin particle line (Fig. \ref{metal.fgr:op} (c)).
The force responsible for the particle migration towards the centre of the capillary is called acoustic radiation force (ARF). The ARF arises from the interactions between the incident acoustic field and the acoustic field scattered at a particle. For particles whose radius is smaller than the viscous boundary layer thickness
\begin{equation}
    \delta=\sqrt{\frac{\eta}{\pi \rho_0 f}},
\end{equation}
with the density of the fluid at equilibrium $\rho_0$ and the dynamic viscosity $\eta$, viscous effects need to be considered when computing the ARF, since these could even lead to an inversion of the stable particle positions.\cite{10.1103/PhysRevE.100.061102,10.1098/rspa.1994.0150} Here, we deal with particle radii ($r$) just larger than the viscous boundary layer thickness, namely $r \sim \SI{0.58}{\micro\meter} > \delta \sim \SI{0.44}{\micro\meter}$, while the acoustic wavelength ($\lambda$) is much bigger than the particle radius, meaning that the inviscid ARF theory still offers a good approximation.
The ARF is, therefore, given as \cite{10.1055/s-2004-815600}

\begin{equation}
  \boldsymbol{F}_\mathrm{rad} = -\boldsymbol{\nabla} U,
  \label{metal.eq:Frad}
\end{equation}
with the Gor'kov potential

\begin{equation}
  U=\frac{4}{3} \pi r^3 \left( \frac{1}{2}  \frac{f_1}{ c_0^2 \rho_0 } \langle p_1^2 \rangle - \frac{3}{4} \rho_0 f_2 \langle \boldsymbol{v}_1 \boldsymbol{\cdot} \boldsymbol{v}_1 \rangle \right),
  \label{metal.eq:Gorkov}
\end{equation}
with the incident acoustic pressure field $p_1$, the incident acoustic velocity field $\boldsymbol{v}_1$, the fluid speed of sound $c_0$, and the monopole $f_1$ and the dipole $f_2$ scattering coefficients. $\langle \square \rangle$ denotes time averaging $ \langle \square \rangle = \frac{1}{T} \int_{t_1}^{t_1+T} \square  \,dt$ with any point in time $t_1$ and the period of oscillation $T=1/f$.

Another nonlinear time-averaged effect that needs to be considered in our system is acoustic streaming. The force exerted on particles by acoustic streaming is the Stokes' drag \cite{nyborg1965acoustic,lighthill1978acoustic}
\begin{equation}
  \boldsymbol{F}_\mathrm{str} = 6 \pi \eta r ( \boldsymbol{v}_\mathrm{str} + \boldsymbol{v}_0 - \boldsymbol{v}_\mathrm{prt} ),
  \label{metal.eq:Fstr}
\end{equation}
with the particle velocity $\boldsymbol{v}_\mathrm{prt}$, the background flow $\boldsymbol{v}_0$ and the streaming velocity $\boldsymbol{v}_\mathrm{str}$. The latter is usually determined using numerical methods,\cite{10.1103/PhysRevE.100.061102,10.3390/mi11030240} which allow for the consideration of arbitrary geometry. In our case, we employ the finite element method (FEM) to compute the acoustic streaming field.

Due to the different scaling with the particle radius $r$ of the two forces, namely $\boldsymbol{F}_\mathrm{rad}$ (scales with $r^3$, Eq. \ref{metal.eq:Gorkov}) and $\boldsymbol{F}_\mathrm{str}$ (scales with $r$, Eq. \ref{metal.eq:Fstr}), a critical radius $r_\mathrm{c}$, which signals the transition between the ARF-dominated regime to the streaming-dominated regime, can be derived:\cite{10.1039/C2LC21261G}
\begin{equation}
  r_\mathrm{c} = \sqrt{\frac{3\psi}{2\Phi}} \delta ,
  \label{metal.eq:rc}
\end{equation}
{with the geometry dependent factor $\psi =\frac{3}{8}$ for a standing wave parallel to a planar wall \mbox{\cite{10.1039/c2lc40612h}} and the acoustic contrast factor $\Phi$ valid for a 1-dimensional acoustic standing wave, which can be written as:}
\begin{equation}
    \Phi = \frac{1}{3} f_1+\frac{1}{2}f_2 =\frac{1}{3} \left[ \frac{5\Tilde{\rho}-2}{2\Tilde{\rho}+2} - \Tilde{\kappa}\right],
    \label{metal.eq:Phi}
\end{equation}
in which relative compressibility $\Tilde{\kappa} = \frac{\kappa_\mathrm{p}}{\kappa_\mathrm{f}}$ and equilibrium density $\Tilde{\rho} = \frac{\rho_\mathrm{p}}{\rho_\mathrm{f}}$ reflect the ratios between particle $(\square)_\mathrm{p}$ and fluid $(\square)_\mathrm{f}$ properties. In case of PS particles dispersed in water $\Phi \approx 0.17$ and for copper particles dispersed in water $\Phi \approx 0.75$. In commonly used rectangular acoustofluidic channels, {with an excitation frequency of $\sim \SI{1.75}{\mega\hertz}$}, the critical radius evaluates at around $\SI{0.37}{\micro\meter}$ for a copper particle in water and around $\SI{0.78}{\micro\meter}$ for a PS particle in water. Our system, however, appears to have a lower critical radius since particle focusing is feasible with $r \sim \SI{0.5}{\micro\meter}$ for PS particles in water, which might be based on the different channel geometry{, leading to a different $\psi$}.

As mentioned beforehand, the acoustic effects inside the fluid cavity become significant when the system is close to a resonance. The latter can be pinpointed by the analysis of the average acoustic energy density $\left( \overline{E_\mathrm{ac}} \right) $, which is given as\cite{Pierce1991}
\begin{equation}
  \overline{E_\mathrm{ac}} = \frac{1}{V} \int_{V} \left( \frac{1}{2} \rho_0 \langle \boldsymbol{v}_1 \boldsymbol{\cdot} \boldsymbol{v}_1 \rangle + \frac{1}{2} \kappa\langle p_1^2 \rangle  \right)\,dV,
  \label{metal.eq:Eac}
\end{equation}
with the fluid's compressibility $\kappa$ and volume $V$.

\vspace{-0.5 cm}
\section{Materials and Methods}\label{sec:metal_MM}
\subsection{Device Fabrication}
Two piezoelectric elements (10 mm length, 2 mm width, 1 mm thickness, Pz26, Meggitt Ferroperm, Denmark) were glued to the bottom {and side} of a round glass capillary ($\SI{0.5}{\milli\metre}$ {inner diameter}, $\SI{1}{\milli\metre}$ outer diameter, $\SI{76}{\milli\metre}$ length, 1B100-3, World Precision Instruments, Germany) using conductive epoxy (H20E, Epoxy Technology, Switzerland). Copper cables ($\SI{0.15}{\milli\metre}$ diameter) were attached to the piezo with conductive silver paste and glued to the {capillary} with instant glue to increase the mechanical stability.

For the particle ejection experiments, glass nozzles were fabricated by pulling thin-walled borosilicate glass capillaries ($\SI{1}{\milli\metre}$ outer diameter, $\SI{0.75}{\milli\metre}$ inner diameter, TW100-4, World Precision Instruments, Germany) using a pipette puller (P-97, Sutter Instruments, USA) equipped with a $\SI{2.5}{\milli\metre}$ x $\SI{2.5}{\milli\metre}$ platinum/iridium box filament. The nozzle size was tuned by varying the number of pulling cycles, which was achieved by changing the velocity of the puller bars at which the filament heating stops. Nozzle outer diameters in the range from $\SI{10}{\micro\metre} - \SI{50}{\micro\metre}$ were achieved by repetitive pulling cycles until the capillary separates into two identical nozzles. The capillaries' tips (nozzles) were {dipped into} a fluorophilic polymer (Novec™ 1700, 3M™, Switzerland) to render it hydrophobic, enabling water ejection as a thin jet by avoiding wetting of the nozzles' surface, which would eventually lead to droplet formation.

\begin{figure*}[t!]
\centering
  \includegraphics[width=\linewidth]{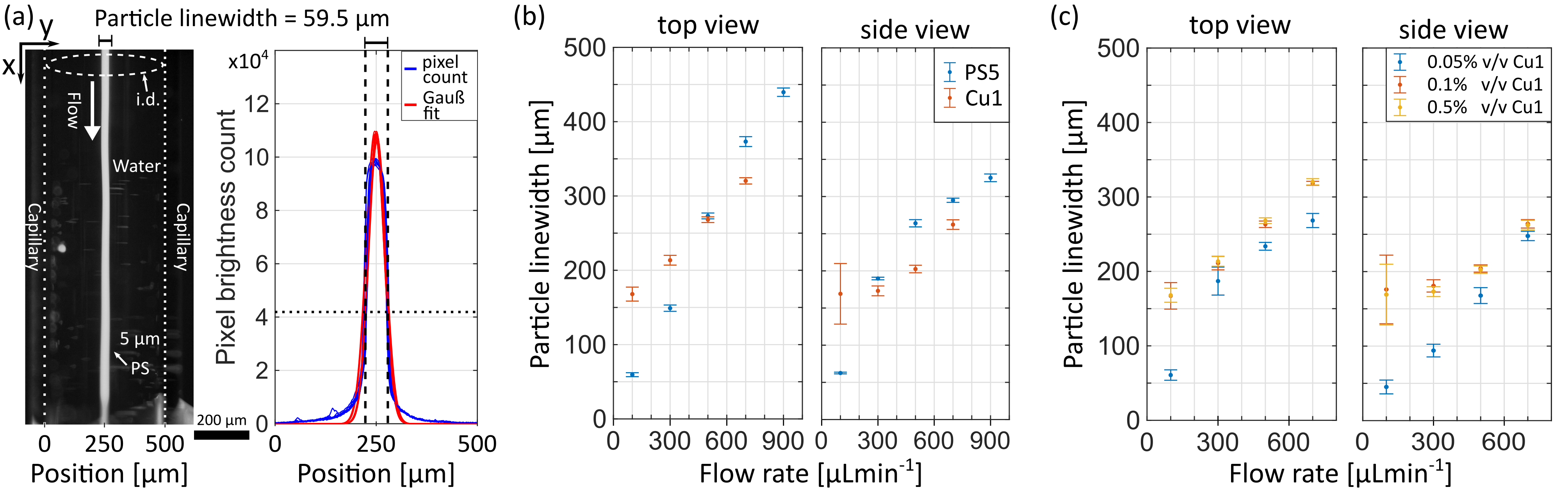}
  \caption{\textbf{Experimental device characterisation.} (a) Optical microscopy picture of PS5 particles diluted in water with $\SI{0,5}{\percent}$ v/v, focused with an excitation frequency $f = \SI{1.74}{\mega\hertz}$ and Voltage $V= \SI{20}{\volt}_\mathrm{PP}$ at a flow rate of $Q = \SI{200}{\micro\liter\per\minute}$. The corresponding analysis of the particle linewidth with the Matlab code described in section \ref{sec:matlab} on the right side. Scale bar corresponds to $\SI{200}{\micro\metre}$ (b) Particle linewidth of PS5 and Cu1 particles diluted in water with a concentration of $\SI{0.5}{\percent}$ v/v. The linewidth increases with increasing flow rate. The linewidth of PS5 and Cu1 particles are comparable despite the difference in size due to the different acoustic contrast factor. (c) Particle linewidth for different initial concentrations of Cu1. The linewidth increases with increasing initial concentration and flow rate.}
  \label{metal.fgr:linewidth}
\end{figure*}
\vspace{-0.5cm}
\subsection{Experimental Setup}
Bulk acoustic wave (BAW) devices are based on the generation of ultrasonic standing waves in a fluid cavity. The waves were coupled into the devices by exciting the piezoelectric elements with an AC signal. The signals were generated \textit{via} a wave generator (AFG-2225, GW INSTEK, Taiwan). The impedance of the piezo varies with its excitation frequency. Since the piezo voltage depends on its impedance, it was verified using an oscilloscope (UTD2025CL, Uni-Trend Technology, China). Water flow inside the capillary was controlled by a syringe pump (neMESYS 290N, Cetoni, Germany). Two syringes were used for the particle ejection experiments; one containing metal particles diluted in distilled water and stabilised with Tween 20 and the other containing solely distilled water. First, the capillary was filled with distilled water. Then, the metal particle flow was slowly increased to achieve the desired concentration and flow rate. If the metal particles were directly inserted into the syringe, the nozzle clogged immediately due to unavoidable particle clumps.
A self-built syringe mixer was utilised to increase the homogeneity of the ejection. The mixer consists of five copper coils linearly arranged along with the syringe, each controlled with a MOSFET (IRF540 n-channel, Infineon, Germany) and a micro-controller (Arduino Nano, Arduino, USA). A cylindrical iron piece was used as a stirrer by moving it up and down inside the syringe. This procedure was essential for copper particles since they sediment much faster due to their higher density and then tend to form big clumps inside the syringe, impeding a proper ejection.

For the optical visualisation, a self-built microscope kit (Cerna®, Thorlabs, Germany) was utilised. A green LED (M505L3, Thorlabs, Germany) excites the particles \textit{via} a dichroic mirror (MD515, Thorlabs, Germany). Before entering the camera (UI-3180CP, IDS Imaging Development Systems GmbH, Germany), the light is filtered by an emission filter (MF535-22, Thorlabs, Germany).
To visualise the side of our glass capillary, individual parts from the company Thorlabs were combined. A camera (UI-3180CP, IDS Imaging Development Systems GmbH, Germany) was connected to a camera tube (WFA4100, Thorlabs, Germany), which is attached to a filter cube housing (WFA2002, Thorlabs, Germany) with a filter cube (MDFM-MF2, Thorlabs, Germany) carrying a dichroic mirror (MD499, Thorlabs, Germany) and an emission filter (MF525-39, Thorlabs, Germany).{ The fluorescent filter sets were removed and two white LEDs were added as background lighting for the top and side view to improve the visibility if metal particles were flown through the capillary.} A high-precision zoom housing (SM1ZM, Thorlabs, Germany), crucial for adjusting the focus, was mounted to the filter cube housing with a combination of three parts (CSA1003, ER1-P4, LCP02/M, Thorlabs, Germany). the particle focusing was observed $\SI{2}{\milli\metre}$ away from the capillaries' exit to ensure a maximal time for the particles to focus. Due to their long working distance, Mitutoyo objectives with a 5x magnification were connected to the microscopes.

For the device characterization, green fluorescent polystyrene (PS) particles (microParticles GmbH, Germany) with $5.19 \pm 0.14 \si{\micro\meter}$ diameter (PS5) and $1.14 \pm \SI{0.04}{\micro\metre}$ diameter (PS1), and copper particles (Nanografi Nanotechnology, Turkey) with $1.16 \pm \SI{0.61}{\micro\metre}$ diameter (Cu1) were used.The initial concentration was achieved by diluting the particles in distilled water. The local concentration after focusing was determined by deriving the ratio of the inner diameter area and the elliptical particle stream area and multiplying it with the initial particle concentration.
\vspace{-0.5cm}
\subsection{Numerical Model}\label{sec:num_model}

A 2D numerical model of the cross-section of the device (including piezo) was built in COMSOL Multiphysics version $5.4$. After studying the device's frequency response, the capillary's displacement as well as the Gor'kov potential and streaming velocity close to the resonance frequency was analysed exposing the highest acoustic energy density in the system $f_\mathrm{res} \approx \SI{1.75}{\mega\Hz}$. Further, particle dynamics were assessed through the streaming velocity and the ARF, which was computed from the Gor'kov potential (Eq. \ref{metal.eq:Frad}).

A user-controlled mesh with several mesh refinements was chosen, especially at the interfaces between different domains, to correctly incorporate streaming into our numerical model. Please refer to Figure S-1 in the Supplemental Material at [URL will be inserted by publisher] for a more detailed mesh analysis. A frequency domain study was used to solve (I) the Thermoviscous Acoustics interface applied to the water domain, (II) the Solid Mechanics interface applied to the capillary, glue and piezo domain, and (III) the Electrostatics interface applied to the piezo domain.
Further, a stationary study of the Creeping Flow interface applied to the water domain was carried out using the acoustic fields from the frequency domain study in the source terms of the streaming equations. This study returns the streaming velocity field as a direct result.
Finally, a time-dependent study of the Particle Tracing for Fluid Flow interface was carried out by taking the acoustic and streaming velocity fields into account.
Please refer to the Supplemental Material SI-1 at [URL will be inserted by publisher] for a more detailed description of the numerical model.
\vspace{-1cm}
\subsection{Determination of the Particle Linewidth}\label{sec:matlab}

A self-written Matlab script was used to analyse the device's performance by estimating the width of the focused particle line. The script takes a video as input and averages 20 frames to account for flow instabilities. After adjusting the region of interest, a concentration profile is created by extracting all pixel brightness values in this region. The brightness values along the flow direction are summed up to receive the brightness distribution along the channel's width. The brightness distribution is fitted with a Gaussian curve and the particle linewidth is determined with the fitted curve by measuring the width of the Gaussian fit after two standard deviations, \textit{i.e.} $\sim \SI{70}{\percent}$ of the particles are within this region (see Fig. \ref{metal.fgr:linewidth}a). The noise was significantly reduced by subtracting a reference picture without any particles from the brightness distribution.

\begin{figure*}[t!]
 \centering
 \includegraphics[width=\linewidth]{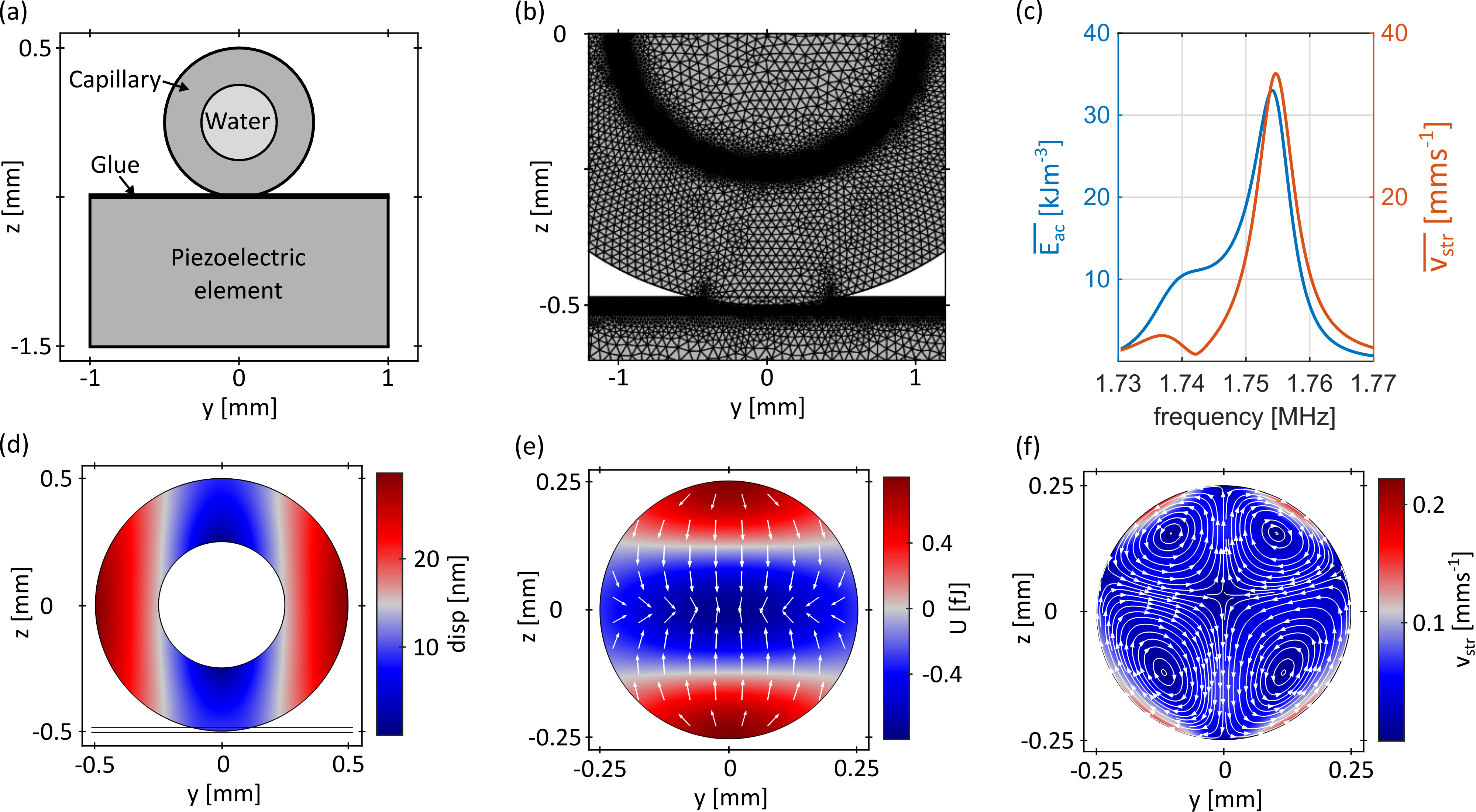}
 \caption{: \textbf{Numerical analysis of the acoustofludic device.} (a) Numerical model of the round capillary with a piezo attached to the bottom \textit{via} a glue layer with $\SI{20}{\micro\metre}$ thickness. (b) Mesh of a small area of the numerical model depicting all domains. (c) Frequency sweep from $1.73$ to $\SI{1.77}{\mega\hertz}$ with steps of $\SI{0.1}{\kilo\hertz}$. At $\SI{1742.1}{\kilo\hertz}$ a minimum in $\overline{v_\mathrm{str}}$ can be found while the $\overline{E_\mathrm{ac}}$ remains at a moderate level. The capillary average displacement magnitude (d), Gor'kov potential with white arrows (logarithmic scaling) as the ARF (e), and streaming patterns (f) at this frequency indicate that particles are pushed towards the centre of the capillary.
}
 \label{metal.fgr:numerical}
\end{figure*}

\section{Results \& Discussion}\label{sec:metal_RD}

\begin{figure*}[ht!]
 \centering
 \includegraphics[width=\linewidth]{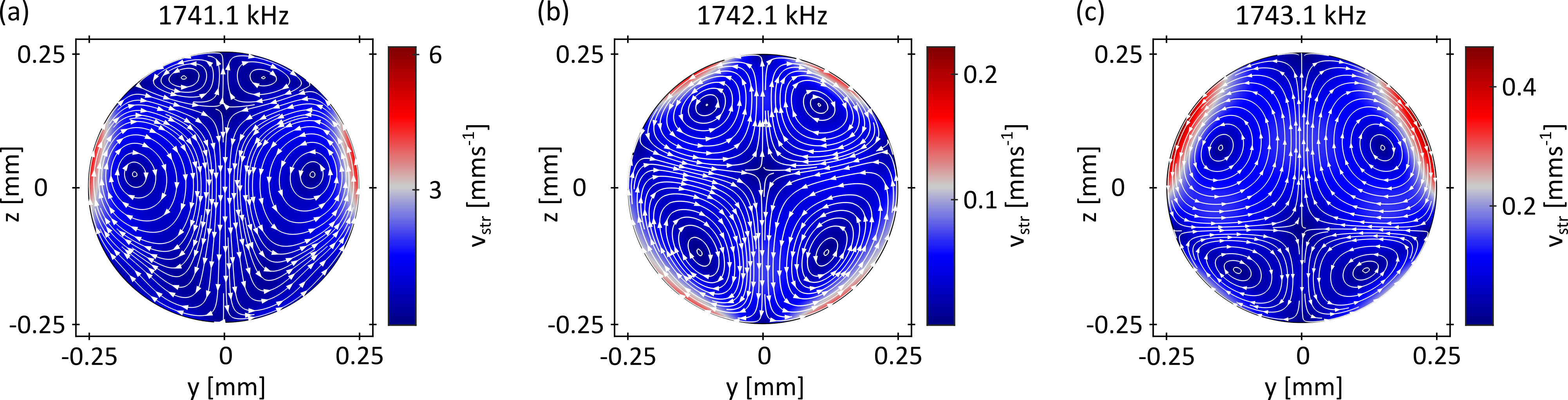}
 \caption{\textbf{Numerical streaming patterns.} Streaming patterns at (a) $f=\SI{1741.1}{\kilo\hertz}$, (b) $f=\SI{1742.1}{\kilo\hertz}$, and (c) $f=\SI{1743.1}{\kilo\hertz}$. The color indicates the amplitude of the streaming velocity, while the arrows on the streamlines indicate the flow direction. The results are given for  $\SI{20}{\volt}_{PP}$ excitation of the piezo. The patterns reveal two transitions; first, from two dominant vortices in (a) to four dominant vortices in (b), and second, from four dominant vortices in (b) to two dominant vortices in (c). In the transition, the two dominant vortices switched from the lower vortices in (a) to the upper vortices in (c). The transitional state in (b) indicates the decrease in the overall maximal streaming velocity, which is beneficial for the manipulation of sub-$\si{\micro\meter}$ particles.}
 \label{metal.fgr:streaming}
\end{figure*}

\subsection{Experimental Device Characterisation}
We experimentally investigated the acoustofluidic device's performance. First, we varied the excitation frequency and chose the frequency that yielded the thinnest particle line by visual inspection. A frequency between $1.67 - \SI{1.78}{\mega\hertz}$ results in the best particle focusing, which is close to the resonance frequency of our numerical model, despite its simplicity (2D geometry). To qualitatively examine the ability of our device to focus particles of different sizes and material properties, we analysed the linewidth of the focused particles by evaluating videos with a self-written Matlab code (Section \ref{sec:matlab}, Fig \ref{metal.fgr:linewidth} (a)). 

We measured the particle linewidth of polystyrene (PS) particles with 
$5.19 \pm \SI{0.14}{\micro\metre}$ (PS5) and $1.14 \pm \SI{0.04}{\micro\metre}$ (PS1) in diameter and copper particles with $1.16 \pm \SI{0.61}{\micro\metre}$ (Cu1) in diameter to characterise the device performance. The particles were dispensed in water with a concentration of $\SI{0.5}{\percent}$ v/v and pumped through the device at flow rates in a range of $\SI{5}{\micro\litre\per\minute}$ {($\SI{0.11}{\milli\metre\per\second}$ average velocity)} to $\SI{900}{\micro\litre\per\minute}$ {($\SI{19.14}{\milli\metre\per\second}$ average velocity)}. Already at $\SI{20}{\volt}_\mathrm{PP}$ the acoustic radiation force in flow direction was strong enough to trap particles close to the piezos below flow rates of $\SI{100}{\micro\litre\per\minute}$ {($\SI{2.13}{\milli\metre\per\second}$ average velocity)}. Therefore, we were not able to analyse the linewidth below a flow rate of $\SI{100}{\micro\litre\per\minute}$ for PS5 and Cu1 particles.
PS1 particles did not get trapped even at low flow rates since the acoustic radiation force is approximately 125 times lower (Eq. \ref{metal.eq:Frad}). Despite the small size, which would usually impede a focus of these particles using acoustic effects, we were able to focus the PS1 particles at a flow rate of $\SI{5}{\micro\litre\per\minute}$ to a line of width $80.6 \pm \SI{6.9}{\micro\metre}$ (top view) and height $65.0 \pm \SI{3.9}{\micro\metre}$ (side view) - see Supplemental Material video 1 at [URL will be inserted by publisher]. The ability to focus below the critical radius might be due to a minimum acoustic streaming velocity, discovered by the numerical analysis (Fig. \ref{metal.fgr:numerical} (c)). At double the flow rate ($\SI{10}{\micro\litre\per\minute}$), the linewidth of the PS1 particles drastically increased to $401.5 \pm \SI{19.9}{\micro\metre}$ (top view) and $277.6 \pm \SI{18.2}{\micro\metre}$ (side view). Above the flow rate of $\SI{10}{\micro\litre\per\minute}$, no particle focusing was observable anymore due to the too short focusing time.

Following the experimental focusing of PS1 particles, we inserted the PS5 particles into the system. At a flow rate of $\SI{100}{\micro\litre\per\minute}$, the PS5 particles could be focused down to a line of width $62.6 \pm \SI{1.1}{\micro\metre}$ (top view) and height $59.5 \pm \SI{2.6}{\micro\metre}$ (side view), corresponding to a local concentration of $\SI{34}{\percent}$ v/v. With a linear increase of the flow rate, the linewidth also increases linearly as the time available to manipulate the particles inside the device is reduced. Even at a flow rate of $\SI{900}{\micro\litre\per\minute}$, we were able to focus the PS5 particles down to a line of width $439.6 \pm \SI{5.7}{\micro\metre}$ (top view) and height $325.2 \pm \SI{5.2}{\micro\metre}$ (side view).

Next, we inserted Cu1 particles into the device. Due to significantly larger density and compressibility difference to water{($\sim 4.4$x bigger acoustic contrast factor (Eq. \mbox{\ref{metal.eq:Phi})})}, copper particles experience a larger ARF compared to the PS particle of the same size. This leads to similar particle linewidths for Cu1 and PS5 particles despite their difference in size. At a flow rate of $\SI{100}{\micro\litre\per\minute}$, the copper particles could be focused down to a line of width $168.0 \pm \SI{9.4}{\micro\metre}$ (top view) and height $169.1 \pm \SI{40.8}{\micro\metre}$ (side view), corresponding to a local concentration of $\SI{4.4}{\percent}$ v/v - see Supplemental Material video 2 at [URL will be inserted by publisher]. Even at high flow rates of $\SI{700}{\micro\litre\per\minute}$, we achieved a line of width $320.4 \pm \SI{4.2}{\micro\metre}$ (top view) and height $262.3 \pm \SI{6.4}{\micro\metre}$ (side view). All particle linewidths for the tested flow rates and particle materials are given in Fig. \ref{metal.fgr:linewidth} (b).

Finally, we evaluated the influence of the initial copper particle concentration on the particle linewidth. We tested three different concentrations ($\SI{0.05}{\percent}$ v/v, $\SI{0.1}{\percent}$ v/v, and $\SI{0.5}{\percent}$ v/v) and found an increase in particle linewidth for increasing flow rate regardless of the concentration (Fig. \ref{metal.fgr:linewidth} (c)). Additionally, the particle linewidth increases for increasing initial particle concentrations, which could be attributed to the maximal particle packing density and inter-particle effects. For an initial concentration of $\SI{0.05}{\percent}$ v/v, we were able to focus the Cu1 particles into a line of width $60.8 \pm \SI{7.0}{\micro\metre}$ (top view) and height $45.2 \pm \SI{9.3}{\micro\metre}$ (side view) at a flow rate of $\SI{100}{\micro\litre\per\minute}$, corresponding to a local concentration of $\SI{4.5}{\percent}$ v/v.

\subsection{Numerical Analysis of the Particle Manipulation}

We utilised the numerical model described in Section \ref{sec:num_model} to study the underlying physical phenomena that lead to two-dimensional particle focusing.
A frequency sweep from $\SI{1}{\mega\hertz}$ to $\SI{2.5}{\mega\hertz}$ ($\SI{0.1}{\kilo\hertz}$ step) revealed a strong resonance of the system at $\SI{1754.1}{\kilo\hertz}$, indicated by a peak in the average acoustic energy density ($\overline{E_\mathrm{ac}}$) and average streaming velocity ($\overline{v_\mathrm{str}}$) (see Fig. \ref{metal.fgr:numerical} (c)). Even though the large acoustic energy density and the Gor'kov potential are favourable for particle focusing in the centre of the capillary, the strong acoustic streaming disturbs the ARF-driven focusing of small particles. We confirmed the undesirable influence of the acoustic streaming by particle trajectory simulations, where the copper particles ($r=\SI{0.5}{\micro\meter}$) end up being carried around with the streaming vortices, which contradicts the experimentally observed results.
 However, at a slightly lower frequency of $f=\SI{1742.1}{\kilo\hertz}$, the $\overline{v_\mathrm{str}}$ is reduced by a factor of $40$ in magnitude in comparison to the $\overline{v_\mathrm{str}}$ at the resonance frequency, while the $\overline{E_\mathrm{ac}}$ is reduced by a factor of $3$ and thus is still relatively high. Since the ARF is proportional to the acoustic energy density, this off-peak frequency would be favourable for the particle focusing.
 At $f=\SI{1742.1}{\kilo\hertz}$, the capillary exhibits displacements in the $\si{\nano\meter}$-range, as shown in Fig. \ref{metal.fgr:numerical} (d). The Gor'kov potential in Fig. \ref{metal.fgr:numerical} (e) with maxima at the top and bottom of the capillary and a minimum in the centre indicates the attraction of particles towards the $z = 0$ plane in the glass capillary. The acoustic radiation force arrows in Fig. \ref{metal.fgr:numerical} (e) point towards the  capillary centre, which confirms the particles' attraction towards the centre of the capillary. The acoustic streaming field consists of 4 vortices that have a joint point of zero velocity close to the centre of the capillary \ref{metal.fgr:numerical}(f), an observation that is beneficial for particle focusing.
 
 The streaming velocity at $f=\SI{1742.1}{\kilo\hertz}$ is low compared to the streaming velocity at the nearby frequencies. This can be attributed to the transition of the streaming pattern that appears to weaken the overall acoustic streaming (see Fig. \ref{metal.fgr:streaming}). Using particle trajectory simulations, we confirmed that the influence of the acoustic streaming at $f=\SI{1742.1}{\kilo\hertz}$ is reduced enough to yield the focusing of $r=\SI{2.5}{\micro\meter}$ PS and $r=\SI{0.5}{\micro\meter}$ copper particles in the centre of the capillary after $\SI{500}{\milli\second}$ (see Supplemental Material Figure S-2 at [URL will be inserted by publisher]). Based on the range of flow rates applied in the experiments, the particles spent between $\SI{1178}{\milli\second}$ and $\SI{131}{\milli\second}$ in the focusing region of two transducers ($\SI{20}{\milli\meter}$), justifying the $\SI{500}{\milli\second}$ observed in the numerical simulations (details in the Supplemental Material SI-2 at [URL will be inserted by publisher]). Considering the three-dimensional acoustic field in the experiments and the related variability in the ARF and the streaming velocity field, the simulations nicely fit our experimental results described in the next section.

\subsection{Focused Metal Particle Ejection}

\begin{figure}[t!]
\centering
  \includegraphics[width=\linewidth]{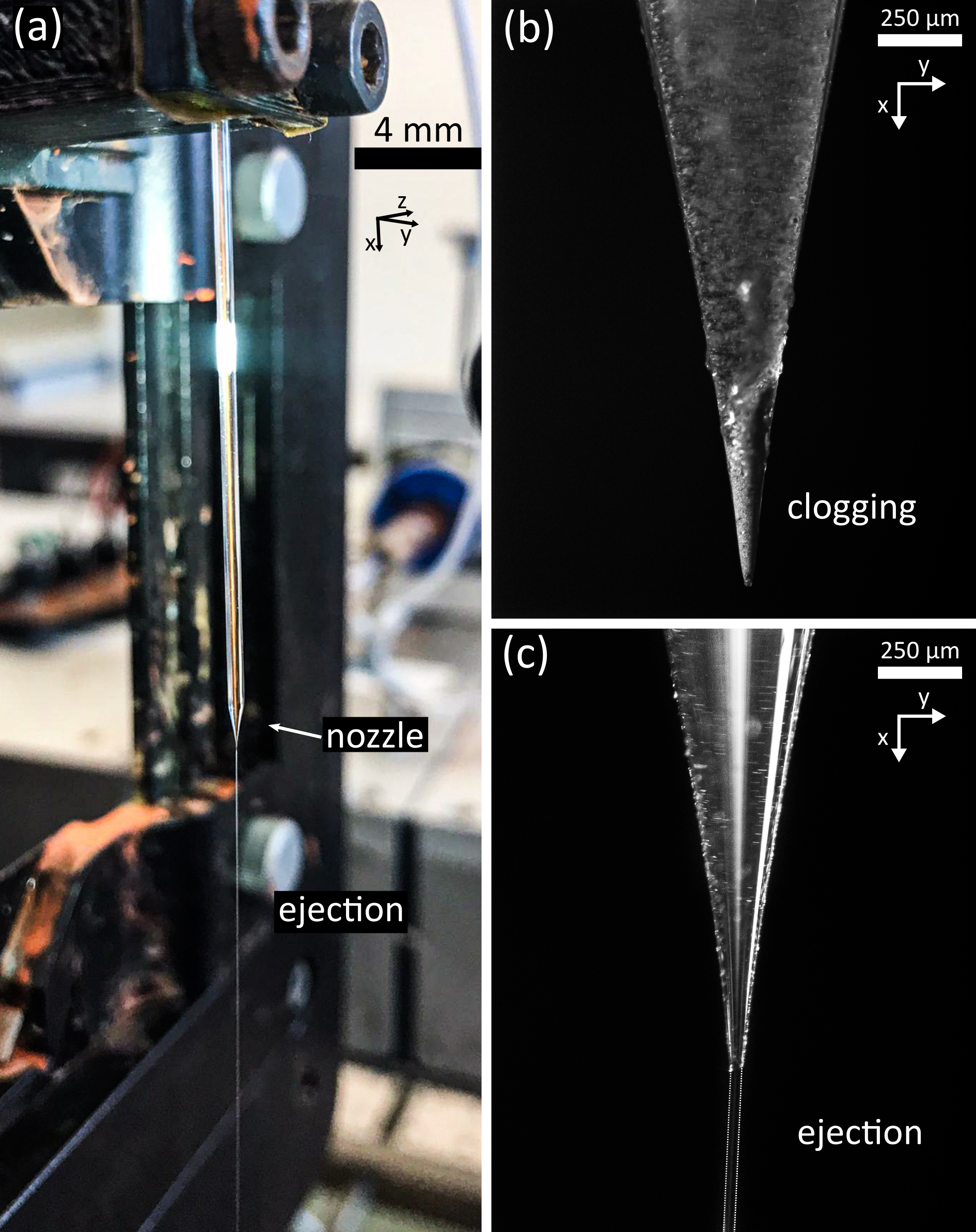}
  \caption{ \textbf{Ejection of focused metal particles through a tapered round capillary.} (a) Photograph of a tapered round glass capillary with a nozzle diameter of $\SI{25}{\micro\metre}$. Water with a flow rate of $\SI{200}{\micro\litre\per\minute}$ is flushed through the system, leading to ejection in the jetting mode. Scale bar corresponds to $\SI{5}{\milli\metre}$. (b) Optical microscopy image of a clogged tapered capillary with a nozzle diameter of $\SI{25}{\micro\meter}$. Cu1 particles with an initial concentration of $\SI{2}{\percent}$ v/v were pumped through the system at a flow rate of $\SI{200}{\micro\litre\per\minute}$ without prefocusing, leading to clogging of the nozzle within seconds. Scale bar corresponds to $\SI{250}{\micro\metre}$. (c) Optical microscopy picture of a tapered capillary with a nozzle diameter of $\SI{25}{\micro\meter}$ while ejecting particles in the jetting mode. Cu1 particles with an initial concentration of $\SI{2}{\percent}$ v/v were flown through the system at a flow rate of $\SI{200}{\micro\litre\per\minute}$. The particles were prefocused within our focusing capillary, which was excited at a frequency of $f = \SI{1.72}{\mega\hertz}$. Scale bars corresponds to $\SI{250}{\micro\metre}$.}
  \label{metal.fgr:ejection}
\end{figure}

For the precise ejection of focused metal particles, as $\textit{e.g.}$ needed for metal 3D printing, a tapered round capillary is connected to the already characterised round glass capillary (particle focusing capillary) with {a combination of silicone and} heat-shrink tubing {(see Supplemental Material Figure S-3 at [URL will be inserted by publisher])}. We tested various nozzle diameters ranging from $\SI{10}{\micro\meter} - \SI{50}{\micro\meter}$. Nozzles with a diameter below $\SI{20}{\micro\meter}$ are clogging even at low initial Cu1 concentrations of $\SI{0.05}{\percent}$ v/v, regardless of prefocusing in the particle focusing capillary. As can be seen in Fig. \ref{metal.fgr:ejection} (b), also nozzles with $\SI{25}{\micro\meter}$ diameter clog within $\SI{10}{\second}$ when Cu1 particles with an initial concentration of $\SI{2}{\percent}$ v/v are flushed through the system at a flow rate of $\SI{200}{\micro\litre\per\minute}$ {($\SI{4.25}{\milli\metre\per\second}$ average velocity)} and the prefocusing system is turned off (see Supplemental Material video 3 at [URL will be inserted by publisher]). However, as shown in Fig. \ref{metal.fgr:ejection} (c), when the ultrasound of our particle focusing capillary is switched on using an excitation frequency of \SI{1.72}{\mega\hertz}, we were able to eject Cu1 particles with an initial concentration of $\SI{2}{\percent}$ v/v at a flow rate of $\SI{200}{\micro\litre\per\minute}$ {($\SI{4.25}{\milli\metre\per\second}$ average velocity)} in a jetting mode (See Supplemental Material video 4 at [URL will be inserted by publisher]). By utilising our prefocusing system, we were able to prevent the clogging of the $\SI{25}{\micro\meter}$ diameter nozzle for $\SI{1}{\milli\litre}$ of sample solution. {The prefocused metal particles are squeezed together in the nozzle resulting in a particle stream of below $\sim \SI{4}{\micro\metre}$ in diameter and ejected with an average velocity of $\sim \SI{1.7}{\metre\per\second}$.} We tested this procedure repeatedly without exchanging the nozzle; hence we believe that much larger volumes can be ejected without clogging the nozzle when utilising our system.

With the combination of our particle focusing capillary and a tapered round capillary, we were able to prove that two-dimensional particle focusing within a round glass capillary can be utilised to impede the clogging of a thin nozzle.

\section{Conclusion \& Outlook}\label{sec:metal_CO}
In this work, we showed for the first time two dimensional focusing of micron-sized particles and their stable ejection through an about twenty times larger nozzle. \newline
We carried out numerical investigations to improve our understanding of two-dimensional particle focusing inside round glass capillaries. We found an excitation frequency at which the influence of the acoustic streaming is weak compared to the acoustic radiation force, explaining our ability to focus particles sized close to the theoretical minimum. \newline
Through experimental evaluations, we were able to determine our device's performance. At a flow rate of $\SI{5}{\micro\litre\per\minute}$, we were able to focus polystyrene particles with $\SI{1}{\micro\metre}$ in diameter into a line with width $80.6 \pm \SI{6.9}{\micro\metre}$ and height $65.0 \pm \SI{3.9}{\micro\metre}$. Additionally, at a $20$ times faster flow speed, we focused five times bigger polystyrene particles into a line of similar size. Copper particles with $\SI{1}{\micro\metre}$ in diameter were observed to behave similarly to $5$ times bigger polystyrene particles due to the significantly higher acoustic contrast.\newline
Finally, we demonstrated the ejection of a $\SI{2}{\percent}$ v/v concentration of $\SI{1}{\micro\metre}$ copper particles through a nozzle of $\SI{25}{\micro\metre}$ diameter, which was unattainable without prior acoustic focusing. With our novel approach, the reliability of systems that rely on particle ejection through a small nozzle and are subject to abrasion could be significantly increased. Therefore, our investigations are expected to be relevant for a wide variety of industrial applications including water jet cutting and metal 3D printing.

\begin{acknowledgments}
The authors would like to express their gratitude for funding by ETH Zurich. This work was funded by the program of the Strategic Focus Area Advanced Manufacturing (SFA-AM), a strategic initiative of the ETH Board. The project can be found under the name "Powder Focusing". We thank Dr. Nino Läubli for his valuable feedback and inputs.
\end{acknowledgments}

\appendix

\section{Data availability statement}\label{sec:dat}
The data that supports the findings of this study are available within the article and its supplementary material. The simulation model is available from the corresponding author upon reasonable request.

\nocite{*}
\section{Notes and references}\label{sec:ref}
\bibliography{rsc}

\end{document}


\newpage
\subsection{Supporting Figures}

\begin{figure}[hbt!]
  \includegraphics[width=1\linewidth]{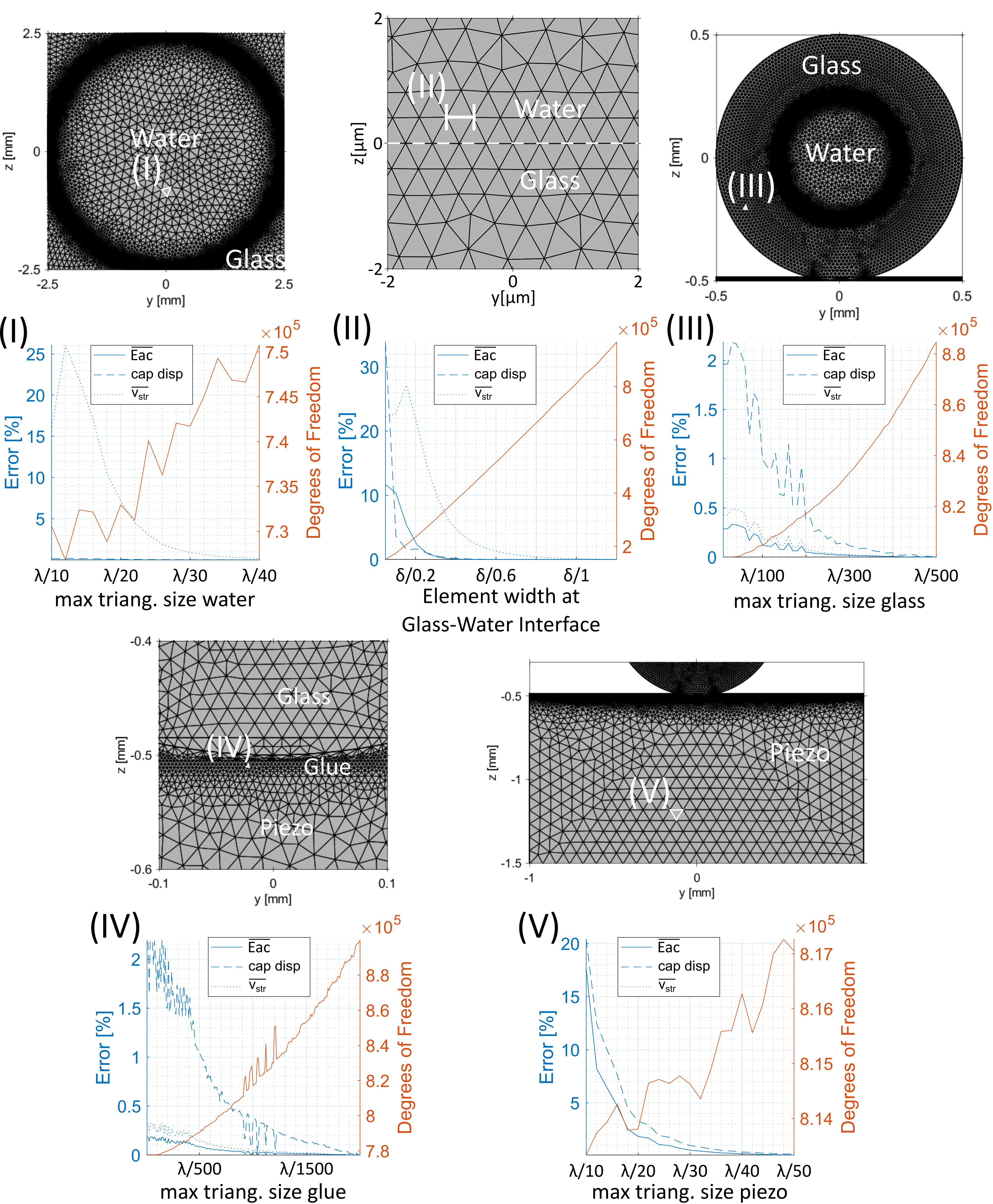}
  \renewcommand{\thefigure}{S-1}
  \label{fgr:mesh}
\end{figure}

\newpage

Figure S-1: \textbf{Mesh study.}  We introduced five mesh parameters reflecting different aspects of the mesh. 
(I) Maximal triangular element size inside the water-domain (max triang. size water) given as fractions of the wavelength ($\lambda/n$). (II) Maximal element width at the boundary between the glass- and water-domain given as fractions of the viscous boundary layer thickness ($\delta/n$). (III) Maximal triangular element size inside the glass-domain (max triang. size glass) given as fractions of the wavelength ($\lambda/n$). (IV) Maximal triangular element size inside the glue-domain (max triang. size glue) given as fractions of the wavelength ($\lambda/n$). (V)  Maximal triangular element size inside the piezo-domain (max triang. size piezo) given as fractions of the wavelength ($\lambda/n$). With $\overline{E_{ac}}$ as the average energy density in the water domain, $\textit{cap disp}$ as the average displacement magnitude of the glass capillary, and $\overline{v_{str}}$ as the average streaming velocity in the water domain. The error is defined as $Error = \left[ 1-(x/x_{ref}) \right]\cdot100$, with $x$ representing one of the evaluated quantities ($\overline{E_{ac}}, \textit{cap disp}, \textit{streaming}$) and $x_{ref}$ representing the value at the finest mesh. We chose the following mesh parameters for the simulations in this work, corresponding to an error below $\SI{1}{\percent}$ with respect to the finest mesh:
max triang. size water $\lambda/40 \approx \SI{21}{\micro\meter} $, Element width at glass-water interface  $\delta/1 \approx \SI{0.4}{\micro\meter}$, max triang. size glass $\lambda/200 \approx \SI{16}{\micro\meter} $, max triang. size glue $\lambda/1000 \approx \SI{3.2}{\micro\meter} $, and max triang. size piezo $\lambda/40 \approx \SI{80}{\micro\meter} $. At a frequency of $f = \SI{1742.1}{\kilo\hertz}$ our mesh had 959891 degrees of freedom. {The distribution of mesh elements was as follows: (I) 1 \% in the Water domain, (II) 80.5 \% in the boundary layer between water and glass, (III) 5.5 \% in the capillary domain, 7 \% in the glue domain, and 6 \% in the piezo domain.} In the Thermoviscous Acoustics interface, we chose quadratic Lagrange elements for the pressure and cubic Lagrange elements for the velocity. In the Creeping Flow interface, we chose P3+P2 for the discretization of the velocity and pressure.
The mesh is generally converged at the chosen mesh parameters. However, if the regions of low streaming velocity in the bulk of the fluid outside resonance are of interest, the element size in the bulk of the fluid (I) needs to be reduced leading to a mesh with 1065620 degrees of freedom at $f = \SI{1742.1}{\kilo\hertz}$.
\newpage

\begin{figure*}[ht!]
 \centering
 \includegraphics[width=\linewidth]{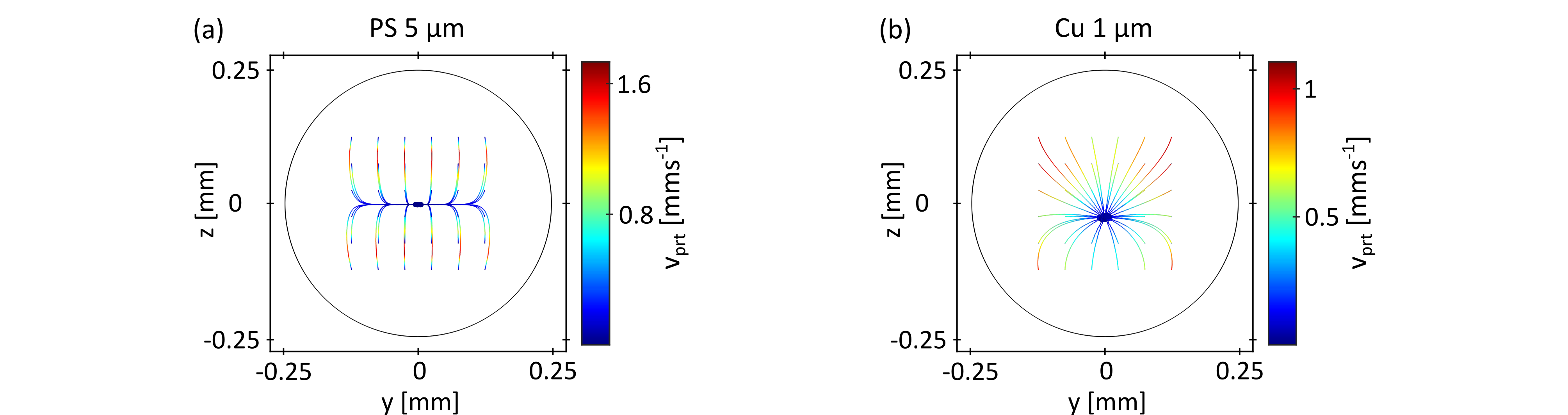}
 \renewcommand{\thefigure}{S-2}
  \caption{\textbf{Numerical particle trajectories.} Particle trajectories estimated at $f=\SI{1742.1}{\kilo\hertz}$ for (a) PS particles with $\SI{5}{\micro\meter}$ in diameter and (b) copper particles with $\SI{1}{\micro\meter}$ in diameter. The $36$ Cu and Ps particles needed approximately $\SI{500}{\milli\second}$ to reach the centre of the capillary. The colour of trajectories indicates the particles speed. Due to a more prominent influence of the acoustic radiation force on the Cu particles, the resulting trajectories point more directly towards the capillary centre. In contrast, the PS particles first move towards the $z=0$ line and then continue to travel towards the centre with a slower velocity.
}
 \label{metal.fgr:trajectories}
\end{figure*}

\newpage

\begin{figure}[ht!]
 \centering
 \includegraphics[width=0.5\linewidth]{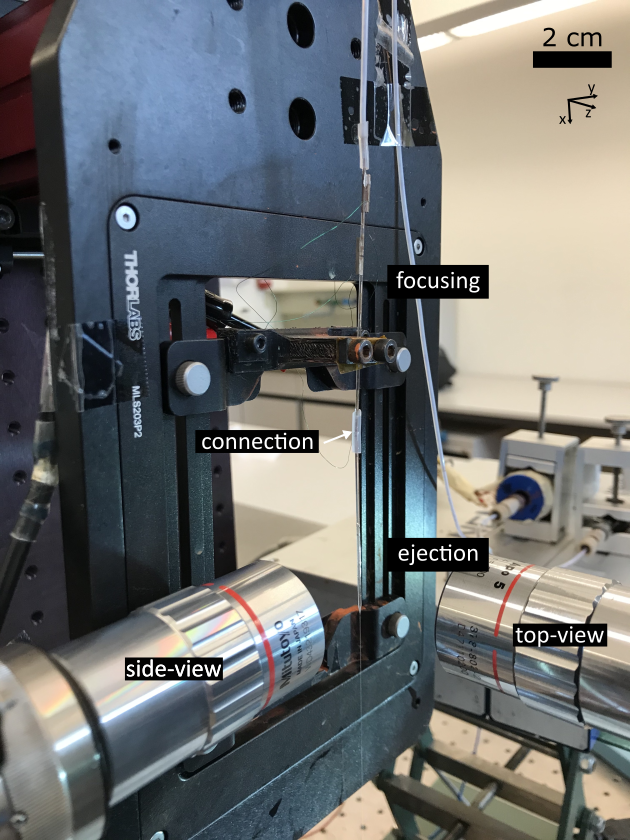}
 \renewcommand{\thefigure}{S-3}
 \caption{\textbf{Photograph of the Setup used for metal particle ejection.} The syringe pumps in the background of the picture are connected to the focusing capillary via PTFE tubing. The ejection capillary is connected to the focusing capillary via silicone and heat-shrink tubing. Scale bar corresponding to 2 cm.}
 \label{metal.fgr:whole_Setup}
\end{figure}

\newpage

\subsection{Supporting Information}

\subsection{SI-1: COMSOL Model}

We used the Solid Mechanics interface to model the solid components of the model (glass and piezoelectric element). To account for damping, we used complex Lamé parameters in combination with the isotropic Linear Elastic Material model. The piezoelectric element was modeled as a Piezoelectric Material. Finally, a $\SI{5}{\micro\metre}$ thin elastic layer was introduced at the interface between the piezoelectric element and glass to model the glue layer (spring constant per unit area). Please refer to Table \ref{metal.tbl:matprop} for all material parameters.

\begin{table}[ht]
\renewcommand{\thetable}{S-1}
  \caption{\ Table of material properties and damping factors.}
  \label{metal.tbl:matprop}
  \begin{tabular*}{1\textwidth}{@{\extracolsep{\fill}}lll}
    \hline
    Parameter & Symbol and value & Unit \\
    \hline
    \textbf{Water at \SI{24}{\degreeCelsius}} \cite{10.1121/1.1913258} \\
    Speed of sound & $c = 1497$ &  {[}\si{\m\per\s}{]} \\
    Density & $\rho^\mathrm{water} = 997$ &  {[}\si{\kg\per\cubic\m}{]} \\\\
    \textbf{Pyrex Glass} \cite{10.1109/t-su.1985.31608} \\
    Density & $\rho^\mathrm{glass} = 2240$ &  {[}\si{\kg\per\cubic\m}{]} \\
    Quality factor & $Q^\mathrm{glass} = 2420$ &  {[}\si{\kg\per\cubic\m}{]} \\
    Lamé parameters & $\lambda = 23.1\times 10^9(1+i/Q^\mathrm{glass})$ & {[}\si{\N\per\square\m}{]} \\
     & $\mu = 24.1\times 10^9(1+i/Q^\mathrm{glass})$ & {[}\si{\N\per\square\m}{]} \\
    \textbf{Piezo Pz26} \cite{Pz26}\\
    Density & $\rho^\mathrm{piezo} = 7700$  & {[}\si{\kg\per\cubic\m}{]} \\
    Electric damping& $tan\delta = 0.003$ & [--]\\
    Quality factor & $Q^\mathrm{piezo} = 100$ & [--]\\
    Stiffness matrix & $c^\mathrm{piezo}_{ij} = 0.965(1+i/Q^\mathrm{piezo})C_{ij}, with$  & \\
    & $C_{11}=C_{22}=155, C_{12}= 94.1$ & {[}\si{\giga\Pa}{]} \\
    & $C_{13}=C_{23}=79.9, C_{33} = 110$ & {[}\si{\giga\Pa}{]} \\
    & $C_{44}=C_{55}=C_{66}=27.3$ & {[}\si{\giga\Pa}{]} \\
    Coupling matrix & $e^\mathrm{piezo}_{ij} = (1+i(1/(2Q^\mathrm{piezo})-tan\delta/2))E_{ij}, with$  & \\
    & $E_{31}=E_{32}= -5.48, E_{33}= 13.6$ & {[}\si{\coulomb\per\square\m}{]} \\
    & $E_{24}=E_{15}= 9.55$ & {[}\si{\coulomb\per\square\m}{]} \\
    Permittivity tensor & $\epsilon^\mathrm{piezo}_{ij} =(1-i(tan\delta))\mathcal{E}_{ij}, with$  & \\
    & $\mathcal{E}_{11}=\mathcal{E}_{22}= 929, \mathcal{E}_{33}= 518$ & {[}\si{\coulomb\per\square\m}{]} \\
    \textbf{Glue layer H20E} \cite{H20E}\\
    Quality factor & $Q^\mathrm{glue} = 10$ & [--]\\
    Spring constants & $k_{11}=2.06(1+i/Q^\mathrm{glue})/5\times10^{-6} $ & {[}\si{\N\per\square\m}{]} \\
    & $k_{22}=8.94(1+i/Q^\mathrm{glue})/5 \times 10^{-6}$ & {[}\si{\N\per\square\m}{]} \\
    \hline
  \end{tabular*}
\end{table}

To account for the piezoelectric effect, the Electrostatics interface was applied to the piezoelectric element and coupled to the Solid Mechanics interface through the Piezoelectric Effect Multiphysics interface. Charge conservation was implemented for the whole domain and an electric potential amplitude of \SI{10}{\volt} was applied to one side of the piezoelectric element, resulting in a \SI{20}{\volt} peak to peak excitation.

The water was modeled with the standard material parameters provided by the software. For computing the acoustic fields inside the capillary, the Thermoviscous Acoustics interface in its adiabatic form was applied to the water domain, and the Thermoviscous Acoustic-Structure Boundary interface was introduced to couple the Solid Mechanics interface and the Thermoviscous Acoustics interface at the water-glass interface.

To simulate the acoustic streaming, we followed a procedure described by Muller et al.\cite{10.1039/C2LC40612H}, resolving the streaming in the viscous boundary layer, as well as in the fluid bulk. We implemented the equations of the acoustic streaming\cite{lighthill1978}
\begin{equation}
    \boldsymbol{\nabla} \langle p_2 \rangle - \eta \nabla^2 \langle \boldsymbol{v}_2 \rangle - \left( \eta_{B} + \frac{\eta}{3} \right) \boldsymbol{\nabla} \left( \boldsymbol{\nabla \cdot} \langle \boldsymbol{v}_2 \rangle \right) = - \rho \boldsymbol{\nabla \cdot} \langle \boldsymbol{v}_1 \boldsymbol{v}_1 \rangle ,
    \label{eq:stream_1}
\end{equation}
\begin{equation}
    \rho \boldsymbol{\nabla \cdot} \langle \boldsymbol{v}_2 \rangle = -  \boldsymbol{\nabla \cdot} \langle \rho_1 \boldsymbol{v}_1 \rangle ,
    \label{eq:stream_2}
\end{equation}
with the time-averaged second-order pressure $\langle p_2 \rangle$, the streaming velocity $\langle \boldsymbol{v}_2 \rangle$, the bulk viscosity of the fluid $\eta_B$, and the acoustic fields that are resolved beforehand, with the Thermoviscous Acoustics interface, namely the acoustic velocity $\boldsymbol{v}_1$ and the acoustic density $\rho_1$. The terms combining the acoustic fields (subscript $1$) in equations \ref{eq:stream_1} and \ref{eq:stream_2} act as source terms for the streaming. In addition, following Baasch et al.\cite{10.1103/PhysRevE.100.061102}, the negative Stokes drift
\begin{equation}
    \boldsymbol{v}_2^{SD} = - \left< \left( \int \boldsymbol{v}_1 \, \text{dt} \boldsymbol{\, \cdot \, \nabla} \right) \boldsymbol{v}_1 \right>
\end{equation}
is applied at the water-glass interface to enforce the no-slip boundary condition at the second order while accounting for the first-order (acoustic) oscillations of the interface.
To implement the described formulation in COMSOL, the Creeping Flow interface was assigned to the water domain. Source terms were enforced through Volume Forces across the domain, while the negative Stokes drift was applied through a Wall boundary condition.
Finally, the second-order pressure field was constrained by setting a Pressure Point Constraint to zero at an arbitrary point of the domain.
\newline
For each simulation, two studies were carried out. First a Frequency Domain study of Solid Mechanics, Thermoviscous Acoustics, Electrostatics, Piezoelectric Effect, and Thermoviscous Acoustic-Structure Boundary interfaces revealed the acoustic response of the system. Subsequently, a Stationary study of the modified Creeping Flow interface was used to find the acoustic streaming inside the water domain while taking the solutions of the Frequency Domain study into account.

\subsection{SI-2: Particle trajectories}
Particle trajectories were simulated within COMSOL, employing the Particle Tracing for Fluid Flow interface. Particles were specified through density and diameter. Two forces were taken into account in calculation of particle trajectories, namely, the Stokes drag
\begin{equation}
  \boldsymbol{F}_{str} = 6 \pi \eta r ( \langle \boldsymbol{v}_2 \rangle - \boldsymbol{v}_{prt} ),
  \label{eq:Fstream}
\end{equation}
due to the particle velocity $\boldsymbol{v}_{prt}$ relative to the acoustic streaming velocity $\langle \boldsymbol{v}_2 \rangle$, and the acoustic radiation force
\begin{equation}
  \boldsymbol{F}_{rad} = - \boldsymbol{\nabla} U,
  \label{eq:Frad}
\end{equation}
with the Gor'kov potential $U$.\cite{10.1055/s-2004-815600} The streaming velocity and the acoustic radiation force were carried on from the previous studies, while the trajectories themselves were evaluated with a Time Dependent study. Particles were released at $t=\SI{0}{\second}$ from a $6\times6$ grid of particles with $100\times r$ spacing for particles with a radius of $r = \SI{1}{\micro\metre}$, and $20\times r$ spacing for particles with a radius of $r = \SI{2.5}{\micro\metre}$.

The time that an individual particle spent in the focusing region of $\SI{20}{\milli\meter}$, along the two piezoelectric transducers, was estimated assuming a Poiseuille flow in the capillary with a circular cross-section. Under these assumptions, the velocity in the centre of capillary along its axis follows as\cite{bruus2008theoretical}
\begin{equation}
    v_P = \frac{2 Q}{\pi a^2},
\end{equation}
with the inner radius of capillary $a$ and the flow rate $Q$. Based on the minimal and maximal flow rates in experiments, \SI{100}{\micro\liter\per\minute} and \SI{900}{\micro\liter\per\minute}, the velocity in the centre of the capillary follows as $\SI{0.0170}{\meter\per\second}$ and $\SI{0.1528}{\meter\per\second}$, respectively. The maximal and minimal times for particle focusing resulting from these velocities follow as $\SI{1178}{\milli\second}$ and $\SI{131}{\milli\second}$, respectively.

\bibliography{metalFocus_supplementary}